\documentclass[aps,prl,floatfix,twocolumn]{revtex4}

\usepackage{graphics}
\usepackage{dcolumn}
\usepackage{epsfig}

\begin{document}

\title
{Excitons and Many-Electron Effects in the Optical Response of
Single-Walled Boron Nitride Nanotubes}
\author{Cheol-Hwan Park}
\author{Catalin D. Spataru}
\author{Steven G. Louie}
\affiliation{Department of Physics, University of California at Berkeley,
Berkeley, California 94720\\
Materials Sciences Division, Lawrence Berkeley National
Laboratory, Berkeley, California 94720}

\date{\today}

\begin{abstract}
We report first-principles calculations of the effects of quasiparticle
self-energy and electron-hole
interaction on the optical properties of single-walled BN nanotubes.
Excitonic effects are shown to be even more important in BN nanotubes than
in carbon nanotubes. Electron-hole interactions give rise to
complexes of bright (and dark) excitons, which qualitatively alter the optical
response. Excitons with binding energy larger than 2 eV are found in the (8,0)
BN nanotubes. Moreover, unlike the carbon nanotubes, theory predicts that these
exciton states are comprised of coherent supposition of transitions from
several different subband pairs, giving rise to novel behaviors.
\end{abstract}
\maketitle
Boron nitride nanotubes (BNNTs) are isoelectronic to carbon nanotubes (CNTs);
however, their electronic properties are quite different. Whereas carbon
nanotubes are metals or semiconductors with different size bandgaps depending
on diameter and chirality~\cite{Saito}, BN nanotubes are
wide gap insulators~\cite{Rubio,Blase}. Although BNNTs have been synthesized
since 1995~\cite{Chopra}, only recently optical measurement on single-walled
BNNTs has been performed~\cite{Lauret}. Theoretical
calculations~\cite{Ando,Spataru1}, as well as experiments~\cite{Wang,Ma,Ma2},
have shown that excitonic effects dramatically
alter the behavior of the optical response of single-walled CNTs. For the
BNNTs, these effects are expected to be even more important due to the wide
band gap nature of BNNTs.

Experimentally it is found that BN nanotubes favor zigzag structure in
current synthesis processes~\cite{Lee}. Thus, we focus our study
on the zigzag tubes. Our calculations on the (8,0) single-walled BNNT show that,
indeed, many-electron effects lead to the formation of strongly bound
excitons of multi-band character with extraordinarily large binding energies,
which dramatically change its optical absorption spectrum.

To compute the optical response, we use the method of Rohlfing and
Louie~\cite{Rohlfing} in which electron-hole excitations and optical spectra
are calculated from first principles in three steps. First, we treat the
electronic ground state with {\it ab initio} pseudopotential density-functional
theory (DFT)~\cite{Sham}. Second, we obtain the quasiparticle energies
$E_{n{\bf k}}$ within the {\it GW} approximation for the electron self-energy
$\Sigma$~\cite{Hybertsen} by solving the Dyson equation:
\vspace{-0.25cm}
\begin{displaymath}
\vspace{-0.25cm}
\left[ -\frac{\nabla^2}{2} + V_{ion} + V_{Hartree} + \Sigma(E_{n{\bf k}})\right] 
\psi_{n{\bf k}}= 
E_{n{\bf k}}\psi_{n{\bf k}}~.
\end{displaymath}
Finally, we calculate the coupled electron-hole excitation energies and optical
spectrum by solving the Bethe-Salpeter (BS) equation of the two-particle Green's
function~\cite{Rohlfing,Strinati}:
\vspace{-0.25cm}
\begin{displaymath}
\vspace{-0.25cm}
\left(E_{c{\bf k}} - E_{v{\bf k}}\right)A^S_{vc{\bf k}} + 
\sum_{{\bf k'}v'c'} \langle vc{\bf k}|K^{eh}|v'c'{\bf k'}\rangle
A^S_{v'c'{\bf k'}} = 
\Omega^S A^S_{vc{\bf k}}~,
\end{displaymath}
where $A^S_{vc{\bf k}}$ is the exciton amplitude, $K^{eh}$ is the
electron-hole interaction kernel, and $|c{\bf k}\rangle$ and
$|v{\bf k}\rangle$ are the quasielectron and quasihole states, respectively.

The DFT eigenvalues and wave functions were obtained within the local
density approximation (LDA)~\cite{Sham} using a
plane-wave basis~\cite{Ihm} with an energy cutoff of  100 Ry.
{\it ab initio} Troullier-Martins pseudopotentials~\cite{Martins} in the
Kleinman-Bylander form~\cite{KB} were used (with cutoff $r_c$ for boron
and nitrogen of 0.79 \AA\ and 0.63 \AA, respectively).
For convergent results to better than 0.05 eV, up to 32 {\bf k} points in
the one-dimensional Brillouin zone were used
for the {\it GW} calculations and for solving the BS equation.
All calculations were carried out in a supercell geometry with a wall-to-wall
intertube separation of 9.5 \AA\ to mimic isolated tubes, together with a
truncated Coulomb interaction in the radial direction in order to eliminate
unphysical interactions between periodic images on the different tubes.
As shown in Ref.~\cite{Spataru2}, it is important to truncate the Coulomb
interaction because these unphysical interactions would increase the effective
screening in the system and hence reduce both the self-energy correction and
the exciton binding energy.
Because of depolarization effects in nanotubes~\cite{Ajiki}, strong optical
response is only observed for light polarized along the tube axis ($\hat z$).
We consider here only this polarization.

\begin{figure}
\includegraphics[width=0.95\columnwidth]{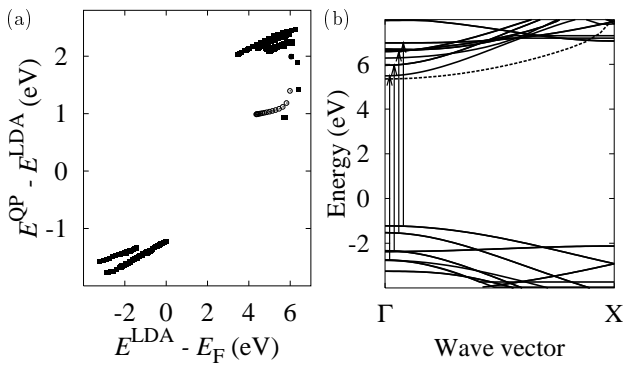}
\caption{Difference between the {\it GW} quasiparticle energy
and the LDA Kohn-Sham eigenvalue plotted as a function
of the energy of the states (a) and quasiparticle band-structure (b) for the
(8,0) SWBNNT. Empty circles in (a) and the dashed line in (b) show the
nearly-free-electron tubule states.}
\label{Fig1}
\end{figure}

\begin{figure}
\includegraphics[width=0.75\columnwidth]{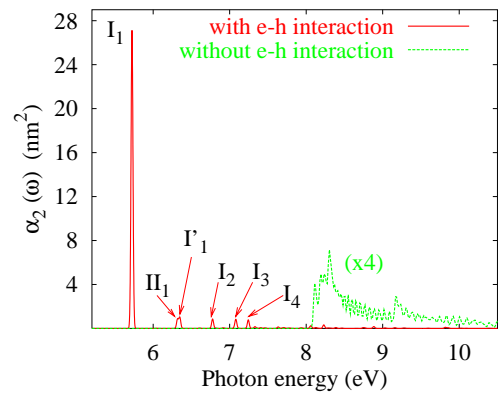}
\caption{(color online). Absorption spectra of the (8,0) SWBNNTs. The imaginary
part of the polarizability per tube $\alpha _2 (\omega)$ is given in unit of nm$^2$.
(See text.) The spectra are broadened with a Gaussian of 0.0125 eV.}
\label{Fig2}
\end{figure}

Figure~\ref{Fig1}(a) shows the quasiparticle energy corrections to the
LDA energy eigenvalues. First, we note that these corrections are quite
large, in comparison to those for bulk hexagonal BN (h-BN) and SWCNTs.
The quasiparticle corrections open the LDA gap of bulk h-BN by $\approx$1.58 eV
near zone center or the {\it $\Gamma$}-point~\cite{Blase_hBN}, while the gap
opening in the (8,0) SWBNNT near the {\it $\Gamma$}-point is $\approx$3.25 eV.
This is a consequence of enhanced Coulomb interaction in reduced
dimension~\cite{Spataru1}. Also, due to its larger gap, which weakens
screening, the quasiparticle  corrections to the gap in the (8,0) SWBNNT are
larger than those for a similar SWCNT (which are $\approx 1.15$ eV near the
{\it $\Gamma$}-point~\cite{Spataru1}). Second, the quasiparticle corrections
have a complex band- and energy-dependence, so for accurate results they cannot
be obtained by a simple scissor shift operation. The corrections depend
on the character of the wavefunction. For example, states of
the fourth lowest conduction band in the LDA bandstructure are localized
inside the tube, and are nearly free-electron-like states.
These tubule states form a separate branch in the quasiparticle
correction diagram with significantly smaller corrections. Figure~\ref{Fig1}(b)
depicts the quasiparticle bandstructure of the (8,0) SWBNNT.
The arrows indicate the optically allowed interband transitions between
four pairs of bands which give rise to the lowest-energy
peak structures in the non-interacting optical spectrum in Fig.~\ref{Fig2}.
(The fourth lowest energy conduction band in the LDA bandstructure becomes
the lowest energy conduction band in the quasiparticle bandstructure.)

Figure~\ref{Fig2} depicts the optical absorption spectrum calculated with and
without electron-hole interaction effects. The plotted quantity is the imaginary
part of the calculated dielectric susceptibility, $\chi =( \epsilon -1)/4 \pi$,
multiplied by the cross-sectional area of the supercell perpendicular to the
tube axis. This quantity $\alpha$, as defined above, gives the polarizability
per single tube in units of nm$^2$; so the susceptibility of an experimental
sample containing a density of $n$ infinitely long tubes per unit area may be
obtained as $\chi = n\alpha$. The absorption profile changes dramatically when
the electron-hole interaction is taken into account. We use the label $I$, $I'$,
and $II$ to denote the excitons that give rise to the absorption peaks in the
figure. Subscripts 1, 2, 3, and 4 refer to the ground, first-excited,
second-excited, and third-excited states of a particular bright exciton series,
respectively. We observe a strong absorption peak at 5.72 eV, which
corresponds to a bound exciton ($I_1$) with a binding energy of 2.3 eV.
The area under this peak, which may be used in comparing with experiment,
is 0.87 nm$^2$eV.
Excitons $I_1$ and $I'_1$ are different states, made up of transitions
from the same set of four pairs of valence and conduction subbands
of the (8,0) BNNT, all of which have similar quasiparticle transition energies
from 8.1 eV to 8.3 eV (See arrows in Fig.~\ref{Fig1}(b)). These transitions
are coupled strongly to each other by the electron-hole interaction to form
the lowest optically active states (the singly-degenerate $I_1$ and
doubly-degenerate $I'_1$).
This behavior is very different from the (8,0) SWCNT in which the exciton states
are composed mainly of transitions between a single pair of quasiparticle bands.

\begin{figure}
\includegraphics[width=0.95\columnwidth]{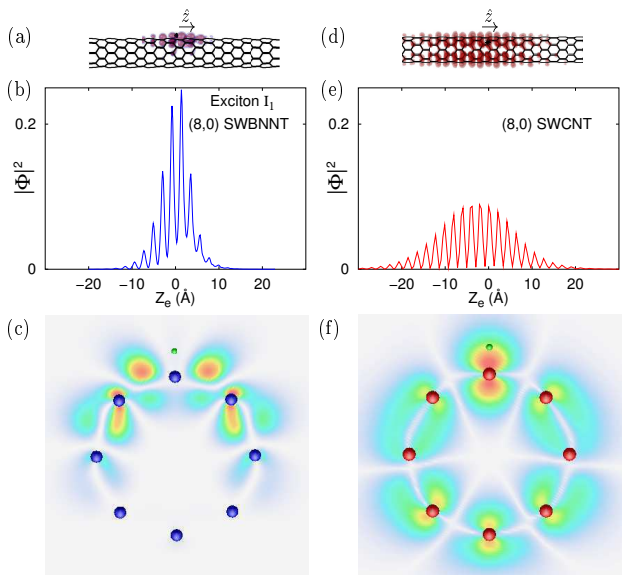}
\caption{(color online). (a)-(c): Wavefunction of the lowest energy bright
exciton of the (8,0) SWBNNT. (a) Isosurface plot of electron probability distribution
$|\Phi({\bf r}_e,{\bf r}_h)|^2$ with the hole fixed at the position indicated
by black star. (b) $|\Phi({\bf r}_e,{\bf r}_h)|^2$ averaged over tube cross section.
Hole position is set at zero. (c) $|\Phi({\bf r}_e,{\bf r}_h)|^2$ evaluated on a
cross-sectional plane of the tube. (d)-(f): Wavefunction of the lowest energy
bright exciton of the (8,0) SWCNT. Plotted quantities are similar to those in (a)-(c).}
\label{Fig3}
\end{figure}

The mixing of transitions of different subbands alters the electron-hole
wavefunction {\LARGE -} localizing further the electron amplitude with respect
to the hole position in real space and making it deviate from a 1-dimensional
behavior with spatial variations in directions perpendicular to the tube axis.
Figure~\ref{Fig3}(a) shows the isosurface plots of the electron
distribution $|\Phi({\bf r}_e,{\bf r}_h)|^2$ with the hole position ${\bf r}_h$
fixed (the black star in the figure) for the first bound exciton ($I_1$).
Figure~\ref{Fig3}(b) quantifies the electron-hole correlation for this state by
plotting $|\Phi|^2$ along the tube axis after integrating out
the electron coordinates in the perpendicular plane (the hole position is set
at zero). The position of the peaks in Fig.\ref{Fig3}(b) corresponds to the
position of plane of boron atoms, i.e., the photoexcited
electron is localized on the boron atoms near the hole. Thus, as expected, the
photo-excitation process corresponds to a transfer of electron from nitrogen
atoms to nearby boron atoms; but the resulting electron and hole amplitudes
are strongly correlated with an extent of only a few inter-atomic distances.
Figure~\ref{Fig3}(c) is a cross-sectional plot, showing the excited electron
probability distribution in a plane that is perpendicular to the tube axis and
contains the hole as well as other nitrogen atoms.

As a comparison to carbon nanotubes,
Fig.~\ref{Fig3}(d)-(f) shows similar quantities as in
Fig.~\ref{Fig3}(a)-(c) but for the first bright bound exciton in the (8,0)
SWCNT~\cite{Spataru1}. In the figure, the hole is fixed
slightly above a carbon atom. The exciton in the (8,0) SWBNNT is significantly
more tightly bound than that in the (8,0) SWCNT and cannot really be viewed
as a 1-D object. The root-mean-square size of the exciton along the tube axis
is 3.67 \AA\ for the (8,0) SWBNNT and 8.59 \AA\ for the (8,0) SWCNT, and their
binding energies are 2.3 eV and 1.0 eV, respectively.
This difference in behavior is due to the wide
bandgap and weaker screening in SWBNNT. Also, we note that while the
binding energy of the excitons in the bulk h-BN is only  0.7
eV~\cite{Arnaud,Wirtz}, the binding energy in the (8,0) SWBNNT is more than
three times larger.

Figures 4(a) and 4(b) show similar quantities as in Fig.~\ref{Fig3}(b)
for the excitons $I'_1$ and $I_2$. For exciton $I'_1$,
the electron is less tightly bound to the hole than in exciton $I_1$.
The state  $I_2$, which is an excited state of exciton $I_1$, is also more
diffuse than $I_1$ and the electron amplitude is not at a maximum near the hole
which is the case for $I_1$(Fig.~\ref{Fig3}(b)).
We also note that, for the (8,0) SWBNNT, there are numerous dark excitons
distributed rather uniformly in energy below and among the bright excitons
shown in Fig.~\ref{Fig2}. The energy of the lowest doubly-degenerate
bound dark exciton is at 4.63 eV. This dark exciton is made up of transitions
from the highest valence band to the lowest conduction band (the NFE tubule
state) in the quasiparticle bandstructure, and has a binding energy of 1.94 eV
with respect to these interband transition energies.

\begin{figure}
\includegraphics[width=0.95\columnwidth]{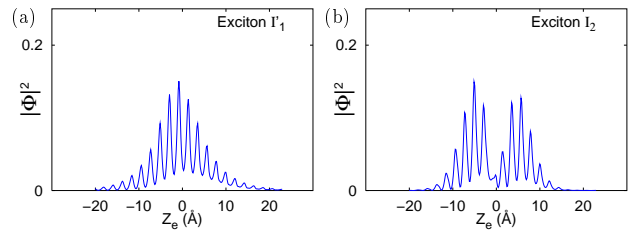}
\caption{(color online). Wavefunctions of excitons of the (8,0) SWBNNT. Plotted quantities are
similar to those in Fig. 3(b).}
\label{Fig4}
\end{figure}

The various lowest-energy exciton states (for both bright and dark excitons)
derived from the various different sets of interband transitions, on the
average, have a large binding energy of about 1.9 eV.
However, the binding energy of the first bright exciton is 2.3 eV.
We ascribe this extra binding energy of about
0.4 eV to the fact that four different sets of interband transitions are
strongly coupled in forming the first bright exciton $I_1$. This strong
coupling mixes states from the different transitions, splits the excitation
energy levels, and increases the binding energy of the final lowest-energy
exciton.

Lauret et. al.~\cite{Lauret} recently measured the optical properties of
samples containing SWBNNTs and observed three absorption peaks at 4.45, 5.5,
and 6.15 eV, respectively. These authors assigned the 4.45 eV peak to originate
from a bound exciton state. In our calculation, the first strong absorption
peak is at 5.72 eV which is about 1.3 eV blue shifted from the observed peak.
A direct comparison between theory and experiment however should not be made.
The average diameter of the tubes measured in the experiment is 1.4 nm while
the diameter of the relaxed (8,0) SWBNNT in our work is 0.65 nm. Also, in the
experiment, there would be tubes with different chiralities.

It is unlikely that the large discrepancy of 1.3 eV is due to
differences in diameter or chirality. At the LDA level, which neglects both
self-energy corrections and excitonic effects, the position of the first broad
absorption peak for SWBNNTs with diameter of 0.48 nm for a (6,0) tube and
2.13 nm for a (27,0) tube is at 4.8 eV and 5.6 eV, respectively~\cite{Guo}.
The difference between them is only 0.8 eV. Furthermore, assuming same
chirality, larger diameter tubes tend to have their first absorption peak
at higher energy. For these reasons, it is difficult to explain the 1.3 eV
discrepancy.
The calculated first peak position for the (8,0) tube is however rather close
to the observed 5.5 eV peak. Also, the observed difference between the second
and the third absorption peak position in the experiment is 0.65 eV, very
close to the difference between the first and the second absorption peaks in
our calculation which is 0.62 eV. But the difference between the first and
the second peak position in the experiment is 1.05 eV.
Another difference is that the theory is for an isolated tube,
while experimentally the tubes are surrounded by a dielectric medium
which can modify the excitation energies. We thus suspect that the observed
second peak at 5.5 eV may be due to an exciton, corresponding in nature to our
first absorption peak at 5.72 eV. However, an additional complication is the
existence of the many dark excitons, which could be made active by
external perturbations. The exact interpretation for the experimental
observations can therefore only be made after further experimental and
theoretical studies.

For SWCNTs, the effect of the surrounding dielectric medium on the optical
spectrum is expected to be small (even though it can be important for the
exciton binding energy) due to an almost cancellation between the
quasiparticle self-energy correction and the binding energy of
excitons~\cite{Spataru1}. For SWBNNTs, screening by external medium would be
more important because intrinsic screening is much weaker. In particular,
in isolated SWBNNTs, we find that the energy difference between the
quasiparticle self-energy correction and the binding energy of the exciton
is large (about 0.9 eV). In the presence of a dielectric medium, we expect
this energy difference to decrease, which would result in a decrease in the
excitation energies.

Among previous theoretical works on the optical properties of BNNTs, Guo and
Lin~\cite{Guo} carried out calculations at the LDA-RPA level without
considering many-electron effects. Their optical absorption spectra are
qualitatively very different from the present final results. From their
results for the (6,0) and (9,0) tubes (showing broad absorption peaks at
4.8 eV and 5.0 eV, respectively), we can deduce a peak position for the
(8,0) tube to be near 4.9 eV, as we find in our LDA-RPA level calculation.
The first peak position in Fig.~\ref{Fig2} with electron-hole interaction
included is blue shifted by about 0.9 eV from that of the LDA-RPA calculation.
Ng and Zhang~\cite{Ng}, on the other hand, used a time-dependent,
localized-density-matrix approach based on a semiempirical Hamiltonian to
compute the optical properties of SWBNNTs. In their work, the absorption
spectrum of (8,0) BNNT has a very broad first peak at an even higher energy than
ours, at near 6.2 eV.

In summary, we have done calculation on the zigzag (8,0) SWBNNT to study the
effects of many-electron interactions on its optical response. The {\it GW}
corrections to the quasiparticle excitation energies of the SWBNNTs
are significantly larger than those for SWCNTs or bulk h-BN. Also, the
quasiparticle energy corrections are found to be complicated so that
interpolation by a simple scissor shift operation is not a good scheme for
accurate calculation. Theory predicts that, unlike the non-interacting
case, the absorption spectrum of the (8,0) SWBNNT is dominated by a huge peak
at 5.72 eV, due to an exciton with a large binding energy of 2.3 eV.
This exciton state is made up of optically-allowed transitions between four
different pairs of subbands. Moreover, an intricate set of dark excitons is
found to exist. Self-energy and electron-hole interaction effects therefore
are even more important in the optical response of the SWBNNTs than in the
SWCNTs.

This work was supported by the NSF under Grant No. DMR04-39768, and by the Director,
Office of Science, Office of Basic Energy Sciences, Division of Materials Sciences
and Engineering, U.S. Department of Energy under Contract No. DE-AC03-76SF00098.
Computational resources were provided by NPACI and NERSC.


\begin{thebibliography}{1}
\bibitem{Saito}
R. Saito, G. Dresselhaus, and M.S. Dresselhaus, {\it Physical Properties
of Carbon Nanotubes} (Imperial College, London, 1998).
\bibitem{Rubio}
A. Rubio, J.L. Corkill, and M.L. Cohen, Phys. Rev. B {\bf 49}, R5081 (1994).
\bibitem{Blase}
X. Blase, A. Rubio, S.G. Louie, and M.L. Cohen, Europhys. Lett. {\bf 28}, 335 (1994).
\bibitem{Chopra}
N.G. Chopra, J. Luyken, K. Cherry, V.H. Crespi, M. Cohen, S.G. Louie, and A. Zettl,
Science {\bf 269}, 966 (1995).
\bibitem{Lauret}
J.S. Lauret, R. Arenal, F. Ducastelle, and A. Loiseau, M. Cau, B. Attal-Tretout,
E. Rosencher, and L. Goux-Capes, Phys. Rev. Lett. {\bf 94}, 037405 (2005).
\bibitem{Ando}
T. Ando, J. Phys. Soc. Japan {\bf 66}, 1066 (1996).
\bibitem{Spataru1}
C.D. Spataru, S. Ismail-Beigi, L.X. Benedict and S.G. Louie, Phys. Rev. Lett.
{\bf 92}, 077402 (2004).
\bibitem{Wang}
F. Wang, G. Dukovic, L.E. Brus, and T.F. Heinz, Science {\bf 208}, 838 (2005).
\bibitem{Ma}
Y.-Z. Ma, L. Valkunas, S.L. Dexheimer, S. M. Bachilo, and G. R. Fleming,
Phys. Rev. Lett. {\bf 94}, 157402 (2005).
\bibitem{Ma2}
Y.-Z. Ma, L. Valkunas, S. M. Bachilo, and G. R. Fleming,
J. Phys. Chem. B {\bf 109}, 15671 (2005).
\bibitem{Lee}
R.S. Lee, J. Gavillet, M. Lamy de la Chapelle, A. Loiseau, J.-L. Cochon, D. Pigache,
J. Thibault, and F. Willaime, Phys. Rev. B {\bf 64}, 121405(R) (2001)
\bibitem{Rohlfing}
M. Rohlfing and S.G. Louie, Phys. Rev. B {\bf 62}, 4927 (2000).
\bibitem{Sham}
W. Kohn and L.J. Sham, Phys. Rev. {\bf 140}, A1133 (1965).
\bibitem{Hybertsen}
M.S. Hybertsen and S.G. Louie, Phys. Rev. B {\bf 34}, 5390 (1986).
\bibitem{Strinati}
G. Strinati, Phys. Rev. B {\bf 29} 5718 (1984).
\bibitem{Ihm}
J. Ihm, A. Zunger, and M.L. Cohen, J. Phys. C {\bf 12}, 4409 (1979).
\bibitem{Martins}
N. Troullier and J.L. Martins, Phys. Rev. {\bf 43}, 1993 (1991).
\bibitem{KB}
L. Kleinman and D.M. Bylander, Phys. Rev. Lett. {\bf 48}, 1425 (1982).
\bibitem{Spataru2}
C.D. Spataru, S. Ismail-Beigi, L.X. Benedict and S.G. Louie, Appl. Phys. A
{\bf 78}, 1129 (2004).
\bibitem{Ajiki}
H. Ajiki and T. Ando, Physica B {\bf 201}, 349 (1994).
\bibitem{Blase_hBN}
X. Blase, A. Rubio, S.G. Louie, M.L. Cohen, Phys. Rev. B {\bf 51}, 6868 (1995).
\bibitem{Arnaud}
B. Arnaud, S. Lebegue, P. Rabiller, and M. Alouani, condmat/0503390.
\bibitem{Wirtz}
L. Wirtz, A. Marini, M. Gruning, and A. Rubio, submitted.
\bibitem{Guo}
G.Y. Guo and J.C. Lin, Phys. Rev. B {\bf 71}, 165402 (2005).
\bibitem{Ng}
M.-F. Ng and R.Q. Zhang, Phys. Rev. B {\bf 69}, 115417 (2004).
\end{thebibliography}
\end{document}